\documentclass[sigconf,natbib=false,10pt,dvipsnames,screen]{acmart}

\usepackage{graphicx}
\graphicspath{ {./figures/} }
\usepackage{paralist}
\usepackage{enumitem}
\usepackage{microtype}
\usepackage{listings}
\usepackage{xspace}
\usepackage[font={small}]{caption}

\setcopyright{none}

\RequirePackage[
  datamodel=acmdatamodel,
  style=acmnumeric,
  ]{biblatex}

\newcommand{\code}[1]{{\texttt{{\small #1}}}\xspace}
\newcommand{\codefoot}[1]{{\texttt{{\footnotesize #1}}}\xspace}

\renewcommand\footnotetextcopyrightpermission[1]{}

\def\Snospace~{\S{}}

\newcommand{\sstitle}[1]{\smallskip\noindent\textbf{#1.\/}}

\newcommand{\sititle}[1]{\smallskip\noindent{\it #1:\/}}

\hypersetup{
	pdftoolbar=true,        
	pdfmenubar=true,        
	pdffitwindow=false,     
	pdfstartview={FitH},    
	pdftitle={},    
	pdfauthor={},     
	pdfsubject={},   
	pdfcreator={},   
	pdfproducer={}, 
	pdfkeywords={}, 
	pdfnewwindow=true,      
	colorlinks=true,       
}

\begin{document}

\title[How do users design scientific workflows?]
{How do users design scientific workflows? \\ The Case of Snakemake}


\author[Pohl et al.]{{\Large Sebastian Pohl$^\dagger$, Nourhan Elfaramawy$^\dagger$, Kedi Cao$^{\dagger\sharp}$, Birte Kehr$^\sharp$, and Matthias Weidlich$^\dagger$}\hspace{.4em}}
\affiliation{%
	\vspace{.5em}
	\institution{{\normalsize $^\dagger$Humboldt-Universit\"at zu Berlin, Germany \quad $^\sharp$Leibniz Institute for Immunotherapy, Regensburg, Germany}}
	\streetaddress{}
	\country{}
\institution{{\normalsize \{sebastian.pohl,nourhan.elfaramawy,matthias.weidlich\}@hu-berlin.de, \hspace{.3em}
\{kedi.cao,birte.kehr\}@klinik.uni-regensburg.de}}
	\vspace{1em}
}

\renewcommand{\shortauthors}{}

\begin{abstract}
Scientific workflows automate the analysis of large-scale scientific data,
fostering the reuse of data processing operators as well as the reproducibility and
traceability of analysis results. In exploratory research, however, workflows
are continuously adapted, utilizing a wide range of tools and software libraries,
to test scientific hypotheses. Script-based workflow engines cater to the
required flexibility through direct integration of programming primitives
but lack abstractions for interactive exploration of the workflow design by
a user during workflow execution. To derive requirements for such interactive
workflows, we conduct an empirical study on the use of Snakemake, a popular
Python-based workflow engine. Based on workflows collected from 1602 GitHub
repositories, we present insights on common structures of Snakemake
workflows, as well as the language features typically adopted in their
specification.
\end{abstract}



\pagestyle{plain}

\maketitle

\section{Introduction}

Scientific workflows define series of discrete programs to automate the analysis of large-scale scientific data~\cite{DBLP:journals/ijhpca/DeelmanPACDMPRT18}. Traditionally, models and systems for scientific workflows have been introduced with a focus on the reuse of standardized data processing operators, as well as the reproducibility and traceability of analysis results. Workflow engines such as Kepler~\cite{DBLP:journals/concurrency/LudascherABHJJLTZ06} and Galaxy~\cite{goecks2010galaxy} provide libraries of standard operators, include collaboration features, and facilitate the execution of workflows on various technical infrastructures. However, they only provide limited support for exploratory research, in which workflows are designed to assess scientific hypotheses. Here, workflows are subject to continuous change, and flexible integration of existing tools and software libraries is important. Script-based workflow engines, such as Snakemake~\cite{DBLP:journals/bioinformatics/KosterR18} and Nextflow~\cite{DBLP:journals/bioinformatics/Nextflow}, offer the required flexibility, but still focus on the specification of workflow at design time and lack the abstractions needed to explore workflow design at run time. Although Snakemake enables the integration of Python notebooks, interactions in these notebooks are decoupled from the workflow definition. Hence, users cannot steer the execution of a workflow based on the insights obtained from intermediate or partial results.

In order to design models and systems for interactive workflows, however, we first need to develop an understanding of the \emph{what} and \emph{how} of workflow design in practice:
\begin{enumerate}[]
\item What are common properties of scientific workflows?
\item How are workflows typically specified?
\end{enumerate}
Answers to the first question shed light on the conceptual requirements for a model for interactive workflows, e.g., in terms of actions to apply to a workflow at run-time. The second question, in turn, focuses on the realization of such a model in the context of a particular workflow engine, e.g., in terms of the language constructs that shall be augmented.

Given the importance of the above questions for effective user support, not only for interactive workflows for exploratory research, but for workflow design in general, it is striking that there exist only a few studies that aim at addressing them empirically. Notably, existing work focused on structural properties of traditional workflows based on reusable operators~\cite{AnalysingScientificWorkflows,DBLP:journals/ijdc/LittauerRLMK12} and also derived abstract categorizations of data processing operators and high-level design principles~\cite{empricialanalysis}. Yet, there is a research gap, framed by the need to derive conclusions on the structural properties of script-based workflows, as well as the language features adopted in their specification.

In this work, we set out to study the questions of \emph{what} and \emph{how} in the design of scientific workflows for the case of Snakemake~\cite{DBLP:journals/bioinformatics/KosterR18}, a popular Python-based workflow engine. Our starting point has been the Snakemake Workflow Catalog,\footnote{\url{https://snakemake.GitHub.io/snakemake-workflow-catalog/}} a listing of more than 2,000 public workflows, mostly in bioinformatics, but also spanning other disciplines, such as astrophysics and Earth-scale infrastructure simulation. Based thereon, we have been able to collect workflows from 1602 GitHub repositories to analyze the structure of the graphs derived for execution, as well as the frequency of specific language features utilized within these workflows.

In the remainder, we first review related work (\autoref{sec:related_work}) and provide background information on Snakemake (\autoref{sec:background}). We then elaborate on how we obtained the collection of workflows (\autoref{sec:collection}), before providing initial insights from its analysis (\autoref{sec:analysis}). We conclude with a discussion of our observations (\autoref{sec:discussion}).

\section{Related Work}
\label{sec:related_work}

Scientific workflow systems, such as
Kepler~\cite{KEPLER}, Galaxy~\cite{goecks2010galaxy}, Taverna~\cite{Taverna}, or Pegasus~\cite{Pegasus} provide rich infrastructures and ecosystems to support users in their data-intensive analysis tasks. Recent script-based workflow engines, such as
Snakemake~\cite{DBLP:journals/bioinformatics/KosterR18} and
Nextflow~\cite{DBLP:journals/bioinformatics/Nextflow}, in turn, focus on the
flexible combination of existing tools and software libraries, as often
required in exploratory research. Those may include basic features for
interactions, mostly in terms of Jupyter notebooks. In general, such notebooks
have been advocated as a basis for the realization of interactivity in
scientific workflows, e.g., in the field of visual analytics~\cite{ReuseWF}.

However, empirical analysis of the use of scientific workflow systems is
scarce. Anecdotal evidence on how users design workflows is available in the
form of case studies, e.g., for Kepler workflows in the BioEarth
project~\cite{UseofKepler}. Small collections of workflows have also been
analyzed to derive common performance
characteristics~\cite{DBLP:journals/fgcs/JuveCDBMV13}.

Only a few studies considered large collections of workflows. Notably, around
400 Taverna workflows from \emph{myExperiment} have been analyzed
in terms of their structural
properties~\cite{AnalysingScientificWorkflows}. The majority of operators (57\%)
were implemented directly by the engine, whereas only (14\%) accounted for
dedicated scripts. Therefore, the workflows in this collection represent
relatively standardized data-processing tasks. A similar study based on the
same repository at a later point in time revealed an increase in workflow
complexity~\cite{DBLP:journals/ijdc/LittauerRLMK12}. The authors further
noticed that workflows contained a large number of data transformation
operators needed to integrate existing tools. This observation can be
interpreted as hinting at the increasing use of workflows for less standardized
analysis tasks.

To obtain requirements for an abstract classification of workflows, a
collection of 260 workflows has been investigated in~\cite{empricialanalysis}
with the aim of identifying their commonalities. Based on these workflows from
various scientific domains, the study devised a collection of motifs and
abstract categories for (i) data processing operators (e.g., data retrieval and
data visualization) and (ii) the design of workflows (e.g., the composition of
workflows or manual tasks in workflows).

We conclude that existing studies did not consider
the structural properties of script-based workflows as well as the use of
particular language features. In this work, we address these gaps and provide
insights into the design and characteristics of script-based
workflows.

\section{Background on Snakemake}
\label{sec:background}

\sstitle{Workflow design}
Snakemake is a Python-based workflow engine~\cite{DBLP:journals/bioinformatics/KosterR18}.
In Snakemake, workflows are defined in a so-called {`snakefile'}, which
includes rules that capture the logical operators of data processing. A {rule}
typically defines three main parts: the {input} files, the {output} files, and
the program to derive the output files from the input files. The program
referenced in a Snakemake rule can be any \code{shell} command, a \code{run}
statement with plain Python code, an external \code{script} (Python, R,
Markdown), or a \code{wrapper} for some script defined in Snakemake's internal
repository. This way, Snakemake provides several mechanisms to integrate
existing tools or software libraries in the workflow.

To apply a {rule} to multiple sets of input files, Snakemake offers an
\code{expand} function that produces input file specifications, essentially
deriving all combinations of its arguments. More flexible control is achieved by \code{input functions}, i.e., Python functions to select input files. Such a
function takes a \code{wildcard} as input to guide the selection of files.

To support conditional execution, a {rule} may be declared to be
\code{checkpoint}, which means that \code{input functions} are re-evaluated
whenever a physical job created for the \code{checkpoint} finished execution. A
{rule} may further include execution-related parameters, such as the number of
threads and the amount of memory to use, or a path to a Conda environment. The
latter enables users to specify a unique software environment per rule.

Snakemake offers four different ways to modularize the design of a workflow: An  \code{include} statement enables the separation of a workflow definition into several files. Workflows may also be combined using a statement \code{module}, which facilitates the reuse of rules among workflows. Through a \code{wrapper}, as mentioned above, external scripts may be integrated from a dedicated repository. Finally, the concept of \code{subworkflow} enables the specification of preliminary steps in data processing, i.e., a \code{subworkflow} will run before the parent workflow to prepare the files needed for the execution of the latter.

Moreover, users may store the configuration of a workflow in dedicated files (in JSON or YAML format). They are structured as dictionaries of parameter keys and values that can be accessed through a workflow's global variable
\code{config}.

\sstitle{Workflow execution}
To execute a workflow, Snakemake needs a target {rule}, which is given
explicitly or assumed to correspond to the first {rule} in the snakefile. Based
thereon, Snakemake derives a set of physical jobs by instantiating each logical
{rule} for each set of \code{input} files (specified directly or computed by an
\code{input function}) that is needed to eventually compute the input of the
target {rule}. Constraints on the execution of the workflow are captured by a
directed acyclic graph (DAG), in which the nodes represent physical jobs and
the directed edges model data dependencies.


A key feature of Snakemake is the abstraction provided for the execution of jobs. That is, jobs can be executed locally or using a distributed computing infrastructure. Thus, users may scale up their experiments from a workstation to compute clusters without making any modifications to the workflow.

Moreover, Snakemake provides basic support for user interactions during
workflow execution through Jupyter notebooks. When executing jobs of a {rule}
that is assigned a \code{notebook} statement, a notebook is started and opened
in a web browser. Although this enables users to explore and visualize the
available data files, it does not provide abstractions to steer or adapt the
execution of the workflow. Once the notebook is closed, the execution simply
continues according to the DAG that was constructed initially.

\section{A Collection of Workflows}
\label{sec:collection}

Our analysis is based on the Snakemake Workflow Catalog,
an automatically generated and continuously growing collection of publicly
available Snakemake workflows. Using this catalog, we collected data in two
ways:

\sstitle{Run cloned repositories} At the time of data collection, we have been
able to extract 1602 GitHub repositories from the catalog. We have been able to
clone 1570 of these 1602 repositories. For each cloned repository, we attempted
to run Snakemake with a flag to build the respective DAG (\textit{-{}-dag}) in
its main folder. Our data collection, therefore, was based on the assumption
that each repository corresponds to one workflow and that Snakemake would be
able to detect the main snakefile in the root directory of the repository.

For 362 out of 1602 repositories, Snakemake successfully ran with the above flag. In the other cases, it failed, for example due to missing files or specification errors. If successful, Snakemake returned the DAG of the workflow in the DOT graph description language. Our analysis of workflow structures in the remainder, therefore, is based on a collection of 362 DAGs,  which we were able to construct in this way.

\sstitle{Query GitHub repositories} We also collected data regarding the use of Snakemake's language features in these repositories. To this end, we queried GitHub, collected the source code of Snakemake workflows, and parsed the code line by line to search for specific key words.
This part of our analysis exploits the data for 1431 of the 1602 repositories, i.e., all repositories for which the snakefile could be queried directly at GitHub, i.e., without cloning the repository.

Specifically, our text search starts with the snakefile in the root directory of the repository. To cover as much of the source code as possible, we then also attempted to resolve \code{include} statements recursively. For 3436 out of the 3550 encountered \code{include} statements, such a resolution was possible, i.e., we have been able to parse the referenced file.

\section{Analysis of the Workflows}
\label{sec:analysis}

Our analysis of the workflow collection focused on the two aforementioned questions on the \emph{what} and \emph{how} of workflow design, exploring (i) structural properties of the workflows (\autoref{sec:analysis_structure}), and (ii) the used language features (\autoref{sec:analysis_language}).

\subsection{Structure of the Workflows}
\label{sec:analysis_structure}

To understand the common structure of Snakemake workflows, we study the granularity with which the analysis task is defined and the presence of specific flow patterns.

\begin{figure}[t]
  \centering
  \includegraphics[scale=0.36]{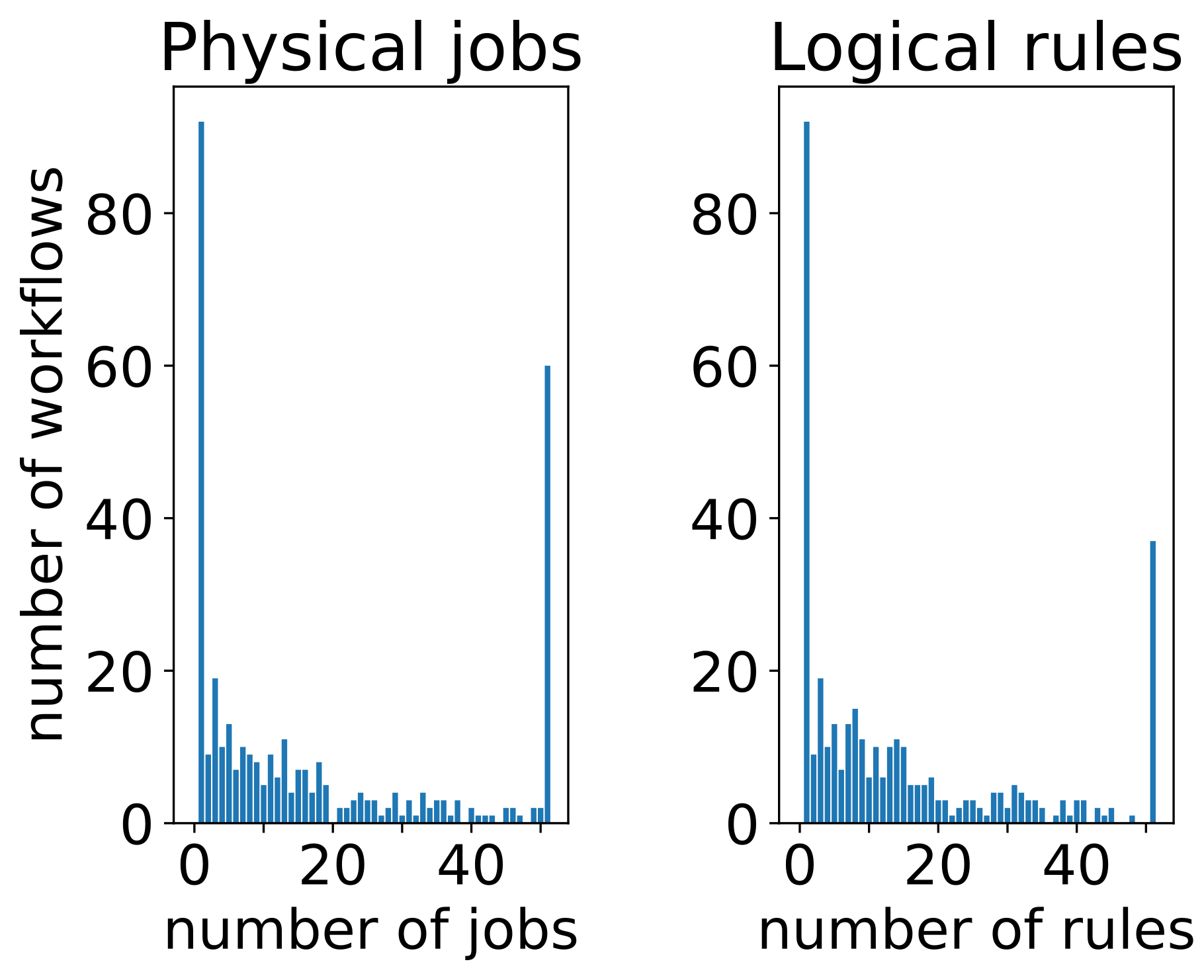}
  \vspace{-.5em}
  \caption{\# Jobs / Rules per workflow (362 DAGs).}
  \label{fig:rules_jobs}
  \vspace{-.5em}
\end{figure}

\begin{figure}[t]
	\centering
	\includegraphics[scale=0.36]{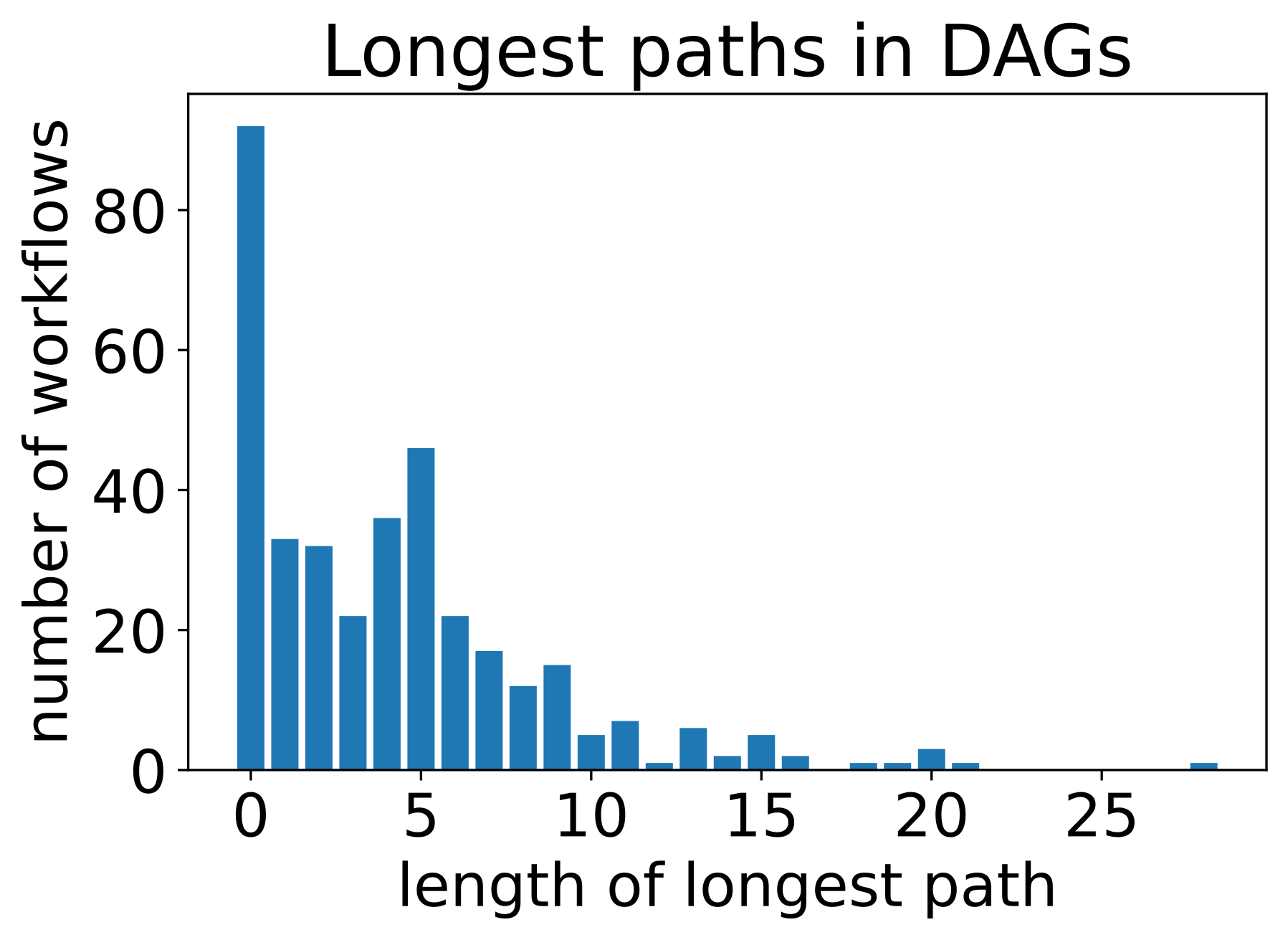}
	\vspace{-.5em}
	\caption{Longest paths in workflows (362 DAGs).}
	\label{fig:sequences}
	\vspace{-.5em}
\end{figure}

\begin{figure*}[t]
  \centering
  \includegraphics[scale=0.32]{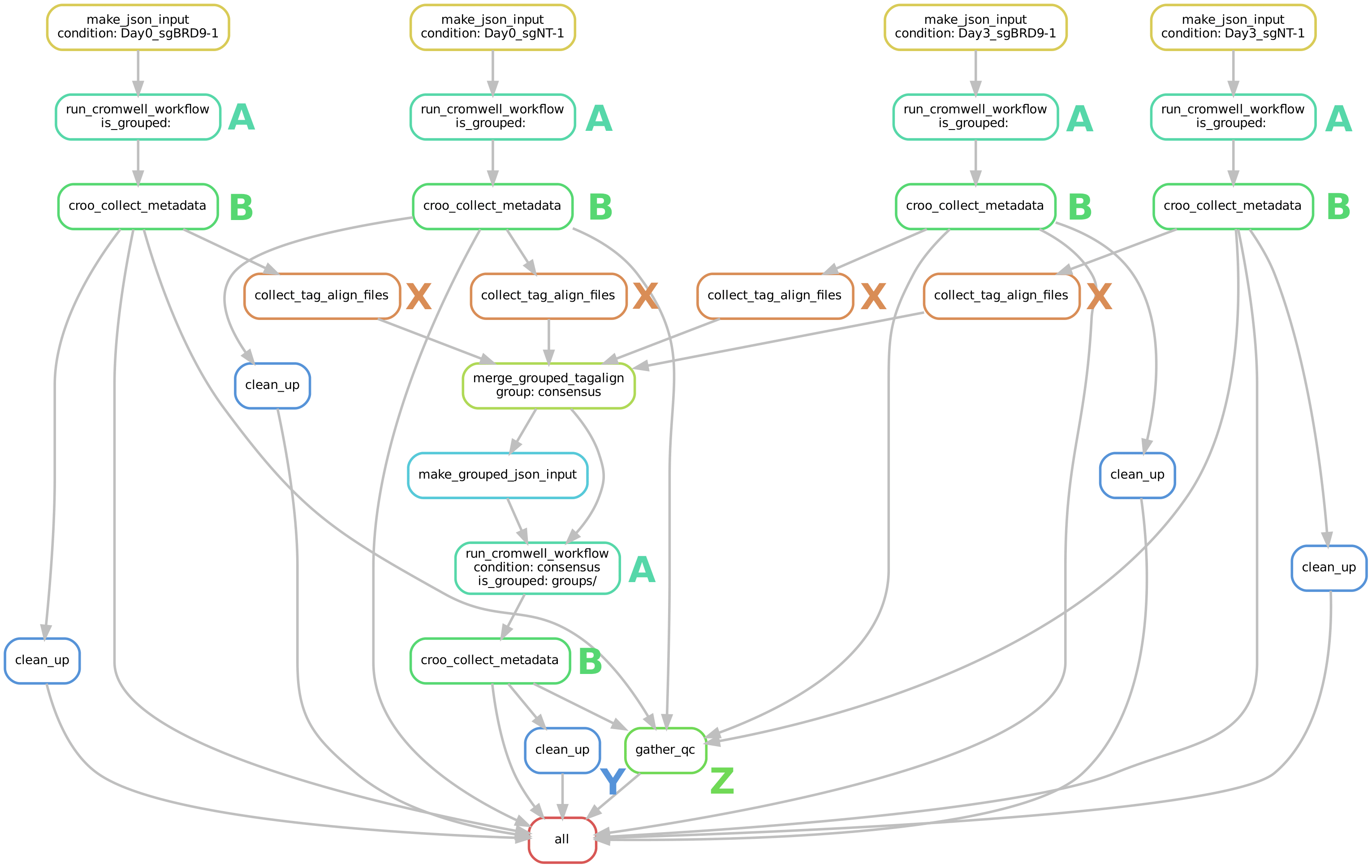}
    \vspace{-.5em}
\caption{DAG of the snakemake workflow from \url{https://GitHub.com/robinmeyers/atac-encode-snakemake}.}
  \label{fig:example}
  \vspace{-.8em}
\end{figure*}

\begin{figure*}[t]
	\begin{lstlisting}[caption=\code{Input function} from the workflow in
		\autoref{fig:example}.,label=fig:input_listing]
def cromwell_inputs(wildcards):
  inputs={'json': os.path.join("jsons",wildcards.is_grouped+wildcards.condition+".json")}
  if (wildcards.is_grouped):
    inputs['tagalign']=os.path.join("results/groups/",wildcards.condition,wildcards.condition+".grouped.tagAlign.gz")
  return inputs\end{lstlisting}
	\vspace{.5em}
\end{figure*}

\sstitle{Granularity}
We start with an assessment of the overall size of the workflows in terms of
the logical {rules} and physical jobs, which provides clues on the granularity
at which the analysis task is specified in a workflow. For the 362
repositories, and hence workflows, for which a DAG could be generated by
Snakemake, the size distributions are given in \autoref{fig:rules_jobs}.
Ignoring a considerable number of degenerated workflows with one {rule}, the
results generally suggest that most workflows comprise
up to 20 rules. However, there exists also a significant number of very large
workflows, with more than 50 rules. Comparing the distributions of logical
rules and physical jobs, there is a notable, but not huge shift, which suggests
that many rules are instantiated only once (as analyzed in more detail later). However, there are also exceptional cases that yield a relatively high number of workflows with more than 50 jobs.

\sstitle{Flow patterns}
Scientific workflows can often be traced back to a few common flow patterns, i.e., sequencing of programs, repetitive behavior, and parallelism.

\sititle{Sequences} The sequentiality in terms of the longest paths of physical jobs in the workflows is illustrated in \autoref{fig:sequences}. Here, the majority of paths is shorter than 10. Combining this result with the sizes of workflows in terms of the total number of physical jobs, see \autoref{fig:rules_jobs}, this hints at a significant amount of jobs that do not have a direct data dependency and, therefore, could be executed independently.

\sititle{Repetitions} To analyze repetitions in workflows, we conducted a depth-first traversal of the DAGs, from the root to all leaf nodes,
and recorded any path that contained at least two physical jobs of the same
logical rule.
We found six DAGs that showed such repetition; only one of them included a
rule that was repeated more than two times.

One example is the workflow in \autoref{fig:example}, in which the rules $A$
and $B$ are apparently applied repeatedly, the first four jobs per rule relate
to individual data samples, whereas the fifth job per rule processed the merged
results of the samples. Interestingly, a closer inspection reveals that only
the first rule is in fact repeated. The second one relies on the \code{input
function} defined in \autoref{fig:input_listing}. It leverages a \code{wildcard},
which is often used to simplify the application of a rule to a large number of
input files, to control the behaviour of rule. That is, depending on the
binding of the \code{wildcard}, the \code{input function} either selects individual
JSON files as inputs or a JSON file of grouped results from an earlier
execution of the rule. As such, the construction is used to introduce
statefulness to this particular rule, illustrating the high degree of
flexibility in script-based workflow specifications.

\begin{figure}[t]
	\centering
	\includegraphics[scale=0.52]{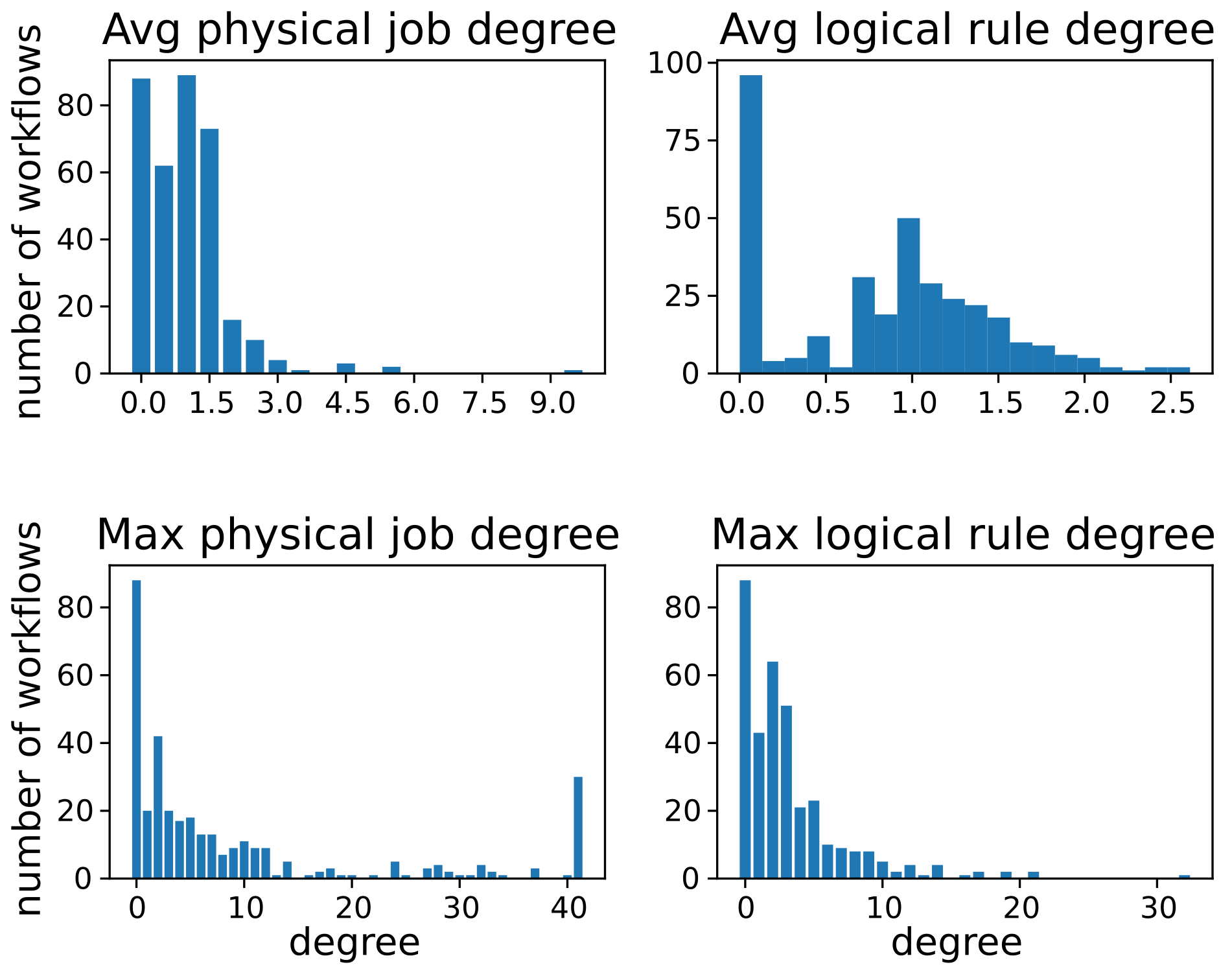}
	\vspace{-.5em}
	\caption{Average and Maximal in-degrees of rules/jobs (362 DAGs), grouping
	degrees >10 (avg) and >40 (max).}
	\label{fig:degree}
	\vspace{-.5em}
\end{figure}

\begin{figure}[t]
	\centering
	\includegraphics[scale=0.3]{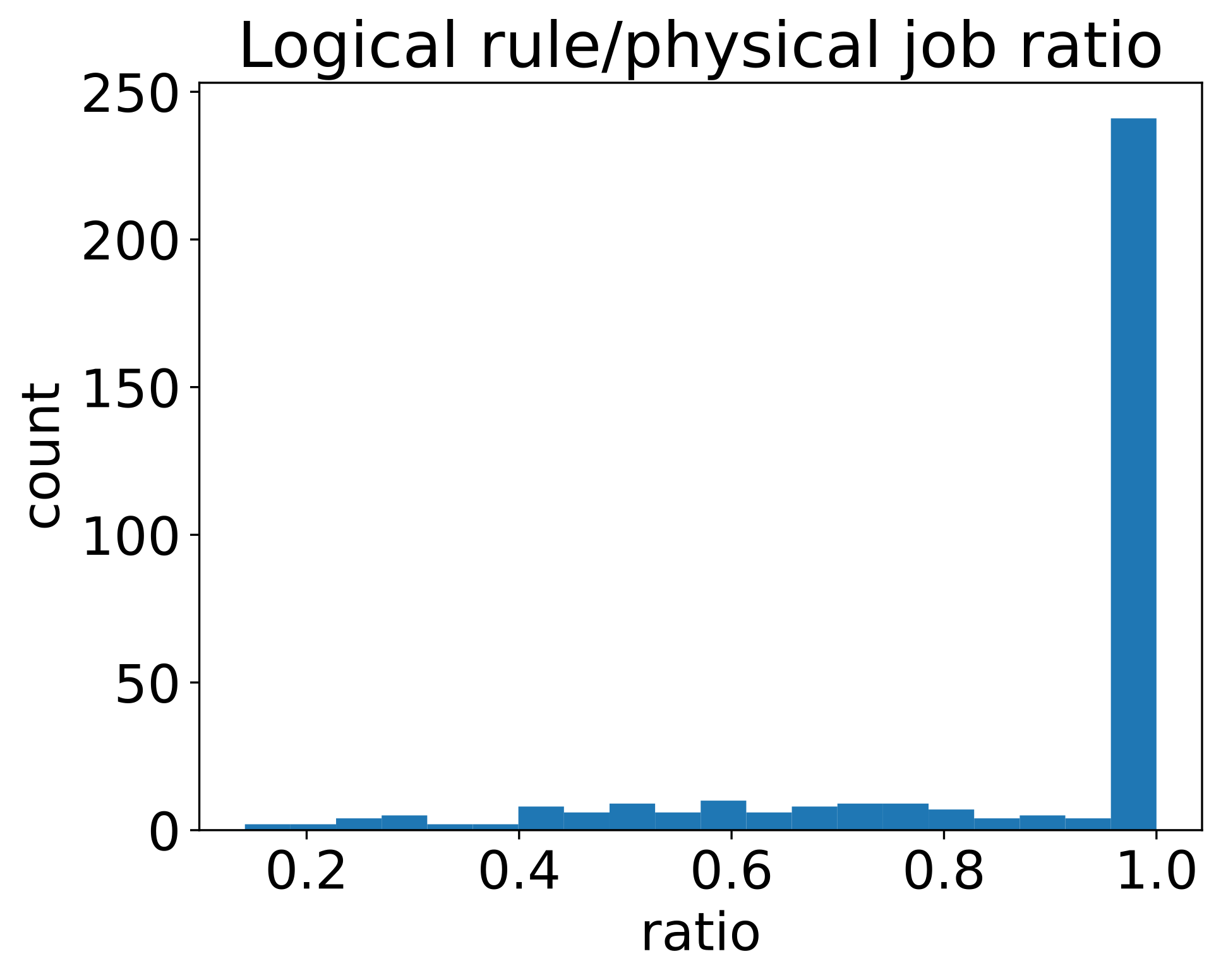}
	\vspace{-.5em}
	\caption{Ratios of rules and jobs (362 DAGs).}
	\label{fig:isolated_ratio}
	\vspace{-.5em}
\end{figure}

\sititle{Parallelism} Physical jobs that are independent may be executed concurrently.
However, such independent jobs may represent data parallelism, in which a
single logical rule is instantiated for various sets of input files, or task
parallelism, in which  certain files are taken as input by multiple different
logical rules. For instance, turning to \autoref{fig:example}, we observe data
parallelism for the jobs of rule $X$ and task parallelism for the jobs of rules
$Y$ and $Z$.

To shed light on the general presence of either type of parallelism in our
workflow collection, \autoref{fig:degree} illustrates the average and maximal
in-degrees of logical rules and physical jobs, respectively. Again, neglecting
the workflows with an average in-degree of zero (mostly degenerated workflows
with a single rule or job), we see that most workflows have an average
in-degree around one for rules and jobs. However, a large number of workflows
also have a maximal in-degree larger than one for rules, hinting at task
parallelism. The distribution for the maximal in-degree for jobs, in turn, is
notably right-shifted. This difference provides evidence of data parallelism, as some workflows have jobs with an in-degree of up to 1440.

\begin{figure}[t]
  \centering
  \includegraphics[scale=0.3]{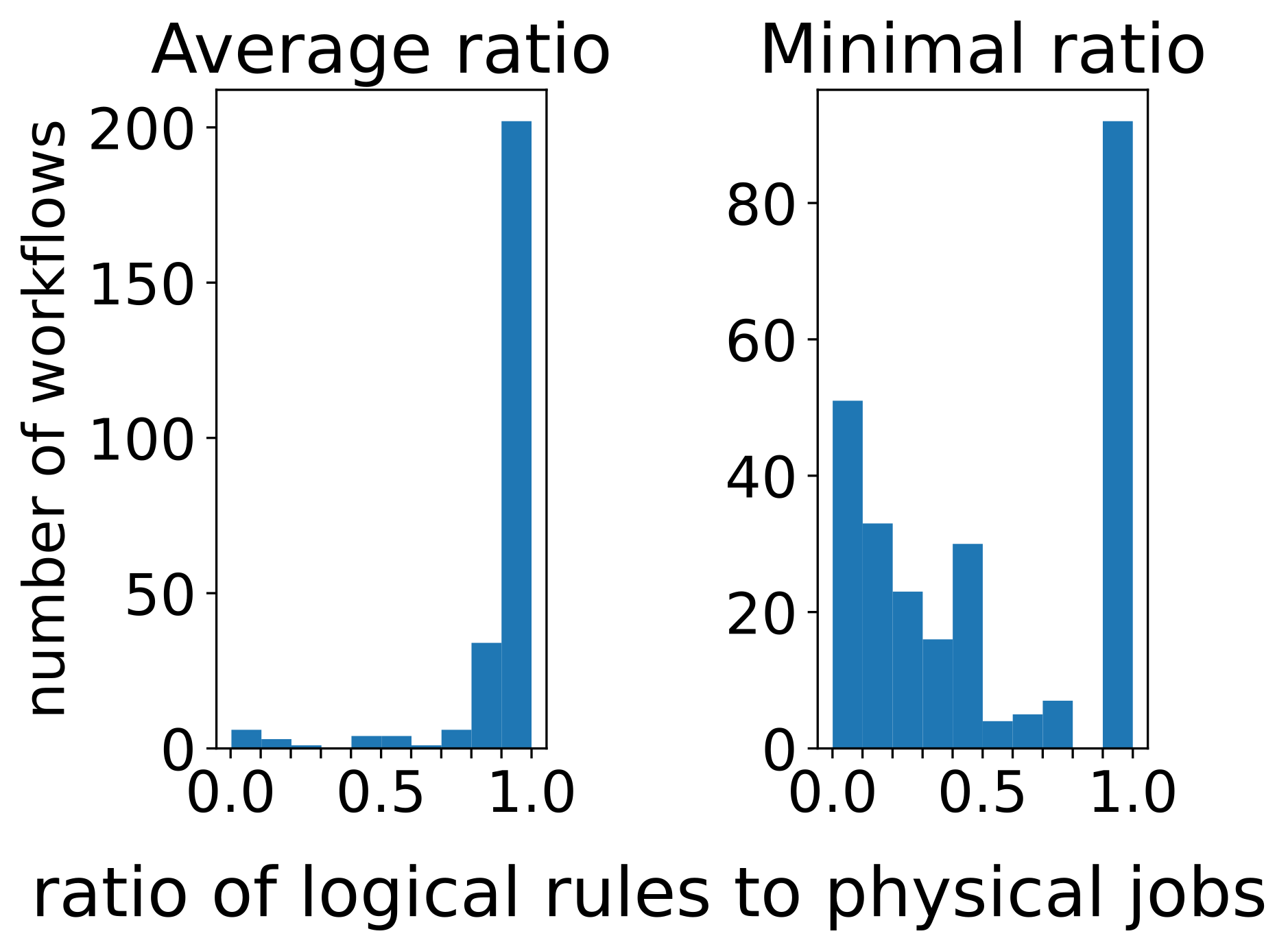}
    \vspace{-.5em}
  \caption{Average and minimal ratio of rules and jobs per workflow (362 DAGs).}
  \label{fig:ratio}
    \vspace{-.5em}
\end{figure}

An alternative view on data parallelism is provided in
\autoref{fig:isolated_ratio}, which shows the ratios of logical rules and their
corresponding physical jobs, over all rules. While a large number of logical
rules are instantiated once, many rules also result in multiple jobs. The
distribution of these ratios over the workflows is illustrated in
\autoref{fig:ratio} in terms of average and minimal values. Here, dozens of
workflows show low minimal ratios, i.e., high data parallelism for at least one
rule.


\subsection{Language Usage}
\label{sec:analysis_language}

Next, we focus on the use of Snakemake language features as observed in the workflow collection.

\sstitle{Modularization primitives}
First, the use of the modularization concepts in Snakemake provides clues on the strength of the coupling of parts of a workflow. For the whole collection of 1431 repositories queried at GitHub, \autoref{tab:modul} illustrates that \code{include} statements and \code{wrappers} are very common, occurring in around a third of the workflows and, typically, many times per workflow. Grouping of rules into \code{modules} is less common, while \code{subworkflows} are rarely used. We interpret these results such that users often manage complexity through modularization within the context of a single workflow, but rarely encapsulate functionality for explicit reuse.

\begin{table}[t]
  \caption{Use of modularization concepts (1431 repos).}
  \label{tab:modul}
  \vspace{-.5em}
  \centering
  \footnotesize
  \begin{tabular}{p{8.2em} r r r r}
  \toprule
  & \textbf{\codefoot{include}} & \textbf{\codefoot{wrapper}} & \textbf{\codefoot{module}} & \textbf{\codefoot{subworkflow}}\\
  \midrule
  Number of repos & 679 & 540 & 200 & 10 \\
  Number of instances & 3550 & 1769 & 415 & 20 \\
  \bottomrule
  \vspace{-.5em}
\end{tabular}
\end{table}

\begin{table}[t]
  \caption{Program types for 17330 rules (1431 repos).}
  \label{tab:execution}
  \vspace{-.5em}
  \centering
  \footnotesize
  \begin{tabular}{ l  r r r r }
  \toprule
  & \textbf{\codefoot{shell}} & \textbf{\codefoot{run}} & \textbf{\codefoot{script}} & \textbf{\codefoot{wrapper}}\\
\midrule
  Number of rules & 13144 & 3426 & 2979 & 1869 \\
  Average number of lines & 5.247 & 9.531 & 2.330 & 2.272 \\
  \bottomrule
\end{tabular}
\end{table}

\begin{table*}[t]
	\caption{Classification of operators in \code{shell} commands into domains,
	including the ratio of rules that contain an operator of a specific domain
	per workflow, averaged over all workflows; the counts of workflows and
	rules that contain operators of a certain domain; and the total number of
	operators per domain (12356 rules, 828 workflows).}
	\label{tab:operator_domains}
	\vspace{-.5em}
	\centering
	\footnotesize
	\begin{tabular}{ r l r r r r l l }
		\toprule
		{ID} & {Operator domain} & {Ratio} & {\# workflow} & {\# rule} &
		{Total} &
		{Description} & {Example operators}\\
		\midrule
		1 & Bioinformatics & 0.326 & 488 & 3928 & 5352 & domain specific tools
		(in our case from bioinformatics) & samtools, bcftools \\
		2 & Control flow & 0.038 & 141 & 579 & 1938 & bash control flow
		keywords & if, for, do \\
		3 & File processing & 0.112 & 314 & 1566 & 3725 & searching, filtering
		and reading files (line by line) & cut, sed, awk, grep \\
		4 & File system & 0.151 & 396 & 1566 & 3725 & organising the file
		system & mkdir, mv, touch, cp \\
		5 & Code execution & 0.157 & 317 & 1566 & 3725 & manage execution
		environments and code execution & conda, python, Rscript \\
		6 & Data handling & 0.086 & 270 & 987 & 1511 & download, pack or unpack
		data & wget, gzip, zcat \\
		7 & Other & 0.568 & 714 & 6959 & 15204 & operator of no clear
		domain/occurred < 10 times total & exec, R, STAR \\
		\bottomrule
	\end{tabular}
\end{table*}

\sstitle{Operators} As mentioned, Snakemake offers various methods to define
the program of a logical rule. To provide a comprehensive overview, we present
the occurrence statistics of the respective keywords in 1431 repositories in
\autoref{tab:execution}. Additionally, we include the average number of lines
in the instructions following these keywords, allowing us to gauge the average
complexity of the rules. In particular, among the 17330 rules analyzed, 5676
rules contained more than one of these keywords, indicating various definitions
of rules.

While the majority of rules primarily define the direct execution of a \code{shell} command, our analysis revealed that all program types are employed to varying degrees.

\begin{figure}[t]
	\centering
	\includegraphics[scale=0.3]{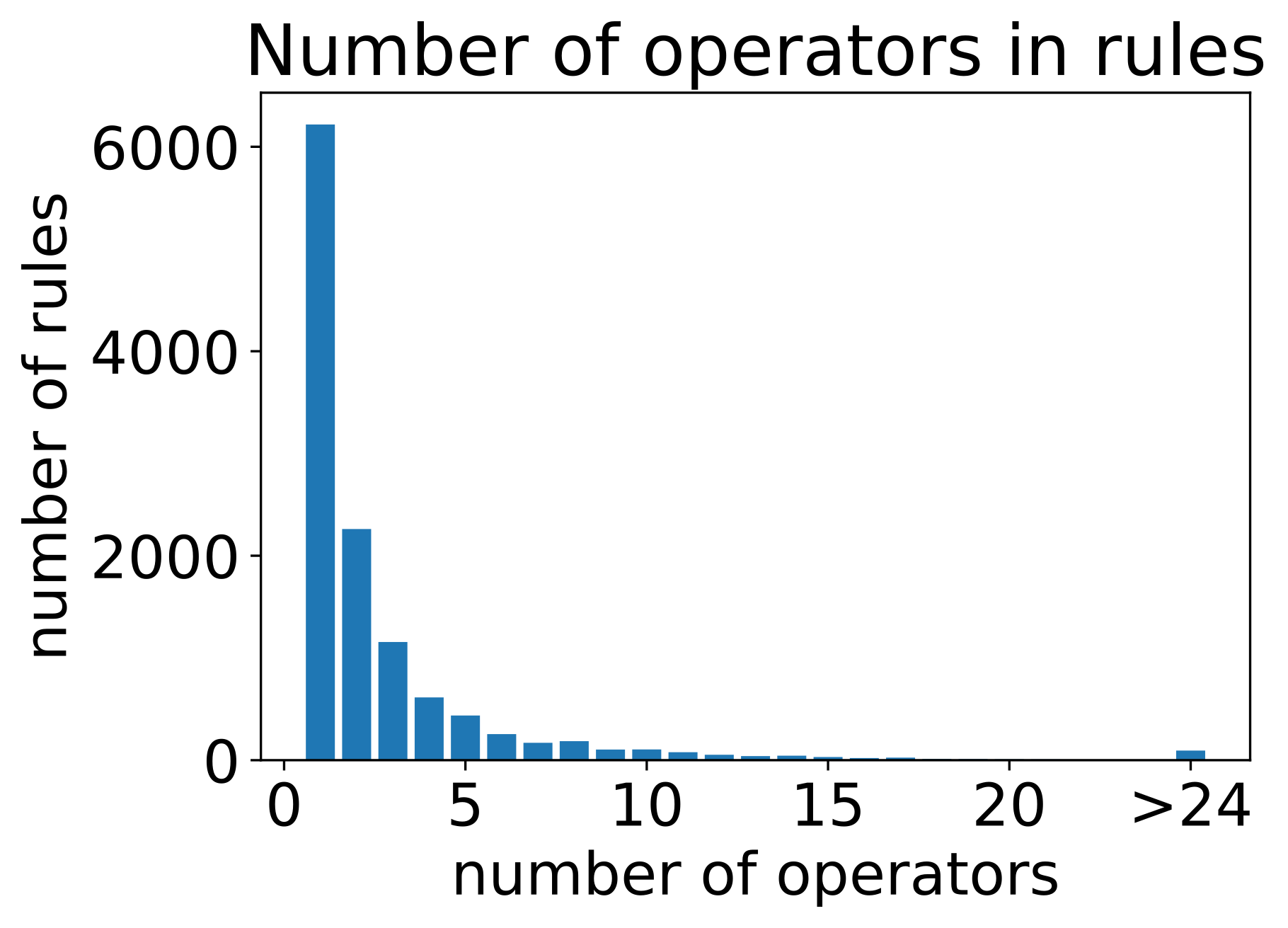}
	\vspace{-.5em}
	\caption{Operator counts (12356 rules, 828 workflows).}
	\label{fig:op_count_dist}
	\vspace{-.5em}
\end{figure}

We further conducted an analysis of the specific operators that are referenced
in \code{shell} commands. By tokenizing the respective command line, we obtain
an understanding of whether the logical rule denotes an atomic program or a
compound program that combines several operators. Starting from the 13144 rules
with a \code{shell} command, tokenization yielded operators for a total of
12356 rules (ignoring, for instance, empty command lines), which are
associated with 828 distinct workflows. The distribution of the number of
operators in \code{shell} commands for these rules is visualized in
\autoref{fig:op_count_dist}. Our results indicate that most rules utilize
only a limited number of operators. Therefore, the \code{shell} commands are
relatively fine-granular, which leaves little opportunity to
further divide them and enable additional parallelism in Snakemake workflows.

Furthermore, we categorized the operators extracted from the \code{shell}
commands of the aforementioned collection into distinct domains. For each
workflow, we
calculated the ratio of rules associated with each domain and derived averages
across all workflows. Additionally, we analyzed the frequency of occurrence of
operators belonging to each domain across the entire dataset. The results are
detailed in \autoref{tab:operator_domains}. A rule is considered part of a
domain if any of the operators associated with that domain are present.
Hence,  a rule can belong to several domains, even all domains.

By examining the results, we first note a significant presence of bioinformatic
operators, which illustrates the popularity of Snakemake in this domain as well
as a high degree of standardization of certain processing steps (as implemented
in tools such as samtools~\cite{li2009sequence}). Moreover, control flow
operators, such as if-statements and for-statements can be found in a significant number of workflows (141 out of 828). These operators hint at conditional branching and
repetition in the workflows and, hence, may point to states in the workflow
that are particularly suited for user interactions (e.g., having a user choose
a branch or determine whether to continue the repetition). Moreover, there is a
considerable presence of operators for file processing and management, generic
code execution, and data handling; as expected in a script-based workflow engine such as Snakemake.

\begin{table}[t]
	\vspace{-.5em}
	\caption{Number of workflows with matching configuration files for operator
		domains (12356 rules, 828 workflows).}
	\label{tab:aggregate_config_matches}
	\vspace{-.5em}
	\centering
	\footnotesize
	\begin{tabular}{ l r r r }
		\toprule
		{Operator domain} & {Total} & {Matched} & Ratio\\
		\midrule
		Bioinformatics & 488 & 115 & 0.136 \\
		\hspace{3mm} \textit{samtools} & \textit{257} & \textit{15} &
		\textit{0.058} \\
		\hspace{3mm} \textit{bcftools} & \textit{109} & \textit{7} &
		\textit{0.064} \\
		\hspace{3mm} \textit{fastqc} & \textit{80} & \textit{11} &
		\textit{0.138} \\
		\hspace{3mm} \textit{bowtie2} & \textit{35} & \textit{11} &
		\textit{0.314} \\
		Control flow & 141 & 59 & 0.418 \\
		File processing & 314 & 67 & 0.213 \\
		File system & 396 & 61 & 0.154 \\
		Code execution & 317 & 20 & 0.063 \\
		Data handling & 270 & 17 & 0.063 \\
		Other & 714 & 113 & 0.1758 \\
		\bottomrule
	\end{tabular}
	\vspace{-2em}
\end{table}

\sstitle{Configuration} Turning to Snakemake's configuration mechanism, we
observe that it is widely utilized, with 712 out of the 1431 repositories
containing a total of 1303 non-empty configuration files. The size of these
configuration files exhibit significant variation, with an average of 64
lines, a median of 32 lines, and a 75th percentile of 71 lines.

Next, we explore whether configuration files are supposedly used to
modify the behaviour of the \code{shell} commands executed within rules.
For each of the aforementioned 828 workflows for which we extracted operators,
we attempted to associate the operators with
the corresponding configuration files. Remarkably, for 236
workflows, i.e., 29\% of the total number, we successfully matched some of the
operators to lines in the configuration files using simple string pattern
matching. This indicates that, indeed, configuration files appear to be
frequently used to fine-tune \code{shell} commands and, hence, rules. A
detailed view on the results of the obtained matches is provided in
\autoref{tab:aggregate_config_matches}. Here, there are notable differences
in ratios of matched occurrences among the domains and operators. For instance,
considering the most frequent
bioinformatics operators, most occurrences of samtools~\cite{li2009sequence}
are not linked to
configuration files (ratio of 0.058), whereas such a link exists for around a
third of the occurrences of bowtie2~\cite{langmead2012fast} (ratio of 0.314).
As such, the use of
bowtie2 is supposedly more frequently adapted as part of the workflow
configuration compared to the use of samtools.

\begin{figure*}[t]
	\begin{lstlisting}[caption=\code{Expand} function in
		rule \textit{tabulate\_sim\_scores} the snakemake workflow from
		\url{https://GitHub.com/gitter-lab/ssps} that
		creates a job with an in-degree of 1440.,label=fig:expand_listing]
		input:
		scores=expand(SCORE_DIR+"/{method}/v={v}_r={r}_a={a}_t={t}_replicate={rep}.json",
		 method=SIM_METHODS,
		v=SIM_GRID["V"], r=SIM_GRID["R"], a=SIM_GRID["A"], t=SIM_GRID["T"],
		rep=SIM_REPLICATES)\end{lstlisting}
	\vspace{.5em}
\end{figure*}


\begin{figure}[t]
	\includegraphics[scale=0.32]{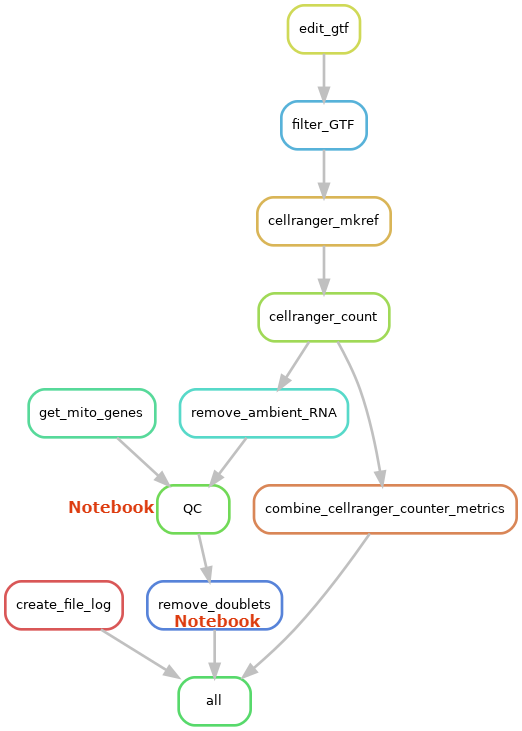}
	\caption{DAG of the snakemake workflow from
		\url{https://GitHub.com/CarolinaPB/single-cell-data-processing}, which
		includes
		Jupyter notebooks.}
	\label{fig:DAG_notebook}
	\vspace{-.8em}
\end{figure}

\sstitle{Dynamic execution} Next, we consider the exploration of language
features for dynamic workflow execution.

\sititle{Expand}
One aspect that stands out is the
widespread use of the \code{expand} function to construct sets of input files.
This method is commonly employed to generate multiple jobs based on various
parameter combinations, as exemplified by the case shown in
\autoref{fig:expand_listing}. Here, a job with an in-degree of 1440 is
created. This particular in-degree value
represents the largest one observed among all the DAGs in
our analysis.
The example illustrates the significant impact of dynamic execution features,
particularly of the \code{expand} function, on the complexity and scale of the
relations between the physical jobs. It also demonstrates the need to
incorporate abstractions to instantiate physical jobs programmatically in order
to handle complex parameter combinations within a workflow in a flexible manner.

\sititle{Input functions \& checkpoints}
When considering the selection of input files through \code{input functions},
our analysis revealed a total of 68 occurrences across 42 of the 1431
repositories. For \code{checkpoints}, we observed 251 occurrences in
94 repositories. While it is commonly expected that input files are directly
selected in most cases, the presence of  \code{input functions} and
\code{checkpoints} in our analysis highlights a certain demand for conditional
execution of workflows.

\sititle{Notebooks}
Additionally, we noticed 47 occurrences of the \code{notebook} statement in 23
of the 1431 repositories. From this observation, it appears that the notebook
feature is not widely used at present, potentially due to its limited
integration in workflow execution.

However, turning to the specific examples that include \code{notebook}
statements, we can derive insights into their purpose. Consider the DAG given in
\autoref{fig:DAG_notebook}, which showcases the integration of two Jupyter
notebooks (as annotated in the figure) within the Snakemake workflow for the
analysis of single-cell RNA data.
The workflow includes two notebooks that serve different purposes: The first
one aims to support quality control by computing various domain-specific
metrics, visualizing the data by means of scatter plots, and then showing the
result of data filtering through further measures and violin plots. The second
one provides a means to remove doublets in single-cell RNA sequence data using
an existing tool. Yet, both steps are explicitly mentioned as points for
exploration, by tuning the filter configuration and computing further measures,
as well as by testing various thresholds in the removal of doublets. As such,
the example illustrates a clear demand for interaction features to explore the
design of a workflow at run-time. Yet, it also indicates the limitations of
existing tooling: If the quality control step indicates an issue, in the
absence of fine-granular control of workflow execution, the whole workflow
needs to be terminated and restarted. Also, the exploration of thresholds for
the removal of doublets is conducted outside the context of the workflow and,
thus, may not influence or steer its execution.

\section{Summary and Discussion}
\label{sec:discussion}

In this study, we embarked on an extensive exploration of a substantial
collection of Snakemake workflows to gain deeper insights into the \emph{what}
and \emph{how} of script-based workflow design. This analysis serves as a
first step towards developing models and methods to support interactive
workflows for exploratory research. In particular, our observations lead to
several conclusions that may help to shape future advancements in Snakemake and
interactive workflow design:

\sititle{(1) Workflow size and opportunities for interaction} Our observations
on
workflow sizes, in terms of physical jobs, logical rules and longest paths,
reveal
that most workflows divide the analysis into sequences of processing steps. In
addition, we also observed that some \code{shell} commands in rules contain
more than one operator. Hence, we conclude that many workflows include various
opportunities for
integrating interactive features that would allow users to assess intermediate
results and, based thereon, adapt the workflow execution as
needed. Also, we consider it remarkable that even though Snakemake does not
provide explicit constructs to implement repetitive behaviour, we observed
workflows that contained loops on the level of physical jobs. Such realization
of repetitive behaviour may also benefit from the integration of user
interactions.

\sititle{(2) Data parallelism and interactions} Many workflows
incorporate data-parallelism, where a single logical rule induces several
physical jobs. This indicates that interaction features to control data
parallelism could be valuable to users. In particular, empowering users to
selectively
evaluate specific parts of the workflow with varying data volumes before
proceeding with full-scale execution can lead to more efficient resource
utilization and reduced processing times. Such evaluation can be expected to be
needed in most cases, as the wide adoption of configuration files provides
evidence for the configurability of many workflows.

\sititle{(3) Language constructs and dynamic workflow steering} Several
observations point to the need
for dynamic workflow steering. The use of \code{input functions} and
\code{checkpoints}, the prevalence of rules with \code{script} program types,
the frequent use of the \code{expand} function, and control-flow statements
(if, for, do) in \code{shell} commands are all examples for this need.
However, Snakemake currently offers limited support for conditional execution.
Incorporating user interactions would offer a richer set of means to adapt and
fine-tune workflow execution, leading to more flexible and adaptable workflows.

\bigskip
\noindent
Overall, our study reveals promising directions for enhancing Snakemake and
exploring novel approaches for interactive workflow design. However, we note
that our workflow collection provides further opportunities to derive insights
on current workflow practice since the repositories include a revision history.
Therefore, in future work, we also intend to analyse the evolution of the
workflows in our collection. Such an analysis can be expected to provide
valuable insights into common change operations and modifications in scientific
workflows. This further exploration will help refine the requirements for
interactive workflows, facilitating the development of tools and features that
cater to the evolving needs of researchers engaged in exploratory data analysis
and computational experimentation.

\vspace{.5em}
\sstitle{Acknowledgments}
Funded by the German Research Foundation (DFG), Project-ID 414984028, SFB 1404
FONDA.

\printbibliography

\end{document}